\documentclass{article}
\usepackage[utf8]{inputenc}
\usepackage{graphicx}
\usepackage{mathabx}

\date{Received June 17 - Accepted August 5, 2020}
\title{The surface brightness of MegaConstellation satellite trails on large telescopes}
\author{Roberto Ragazzoni \\
INAF - Astronomical Observatory of Padova \\}

\begin{document}

\maketitle

\begin{abstract}
    On large telescopes, the trails produced by MegaConstellation satellites will be significantly defocused due to their close proximity. As a result, their apparent surface brightness will be, under a range of conditions, almost constant during their apparent sweep across the focal plane. This paper derives a few simple relationships and evaluate the impact of such trails on operations of large optical ground based facilities.
\end{abstract}

\section{Introduction}

So called MegaConstellation satellites, particularly those on low orbits, have recently raised considerable concern within the professional astronomical community. This paper focuses on their potential disturbance to ground-based optical astronomical observations. It will not address the issues of space debris hazard or potential radio frequency interference with both heavenly human made deep space sources.

More specifically, this paper focuses on the details of the surface brightness in the focal plane of the trails produced by the passage of these satellites. In addition, I will briefly review, offering some point of view, the relationships governing the areal density of these satellites and the likelihood of their being recorded during astronomical observations. A much more thorough estimation of the occurrence of such phenomena derives from statistical methods (Hainaut \& Williams, 2020) and via numerical simulations  (McDowell, 2020). 

Concern about the impact on optical observations has spread, mostly because of the way such satellites reach their intended final orbit. This is achieved through the launch of a relatively large number of satellites into a much lower orbit, followed by continuous, low impulse, maneuvering to the final target altitude (MegaConstellations typically do not use Hohmann maneuvers). As a consequence, the satellites are much brighter and more heavily clustered with respect to their nominal final positions during the initial days and weeks after launch. 

While there are several existing and planned MegaConstellations, this paper uses the StarLink MegaConstellation as reference. I adapted the orbital data from the original FCC requests (FCC Report 2017, later amended by FCC Report 2020a, and a further request in FCC report 2020b), with some degree of approximation and grouping for the sake of readability of the present manuscript. The interested reader should consult the original documents for a detailed description of the MegaConstellation. Note that the number and distribution of satellites considered for the present computations do not include developments in the near future, and an increase by more than an order of magnitude on the overall number of similar satellites can be anticipated (see for example Tyson et al. 2020).

Note also that most of the relationships and findings described in the following are specific to large aperture telescopes. They are not necessarily applicable to modest apertures, such as those used in amateur astrophotography or naked eye observations.

\section{Satellite Numbers}

Satellites in MegaConstellations are usually placed in circular, or almost circular, orbits, characterized by an inclination to the Earth's equator, denoted here by $\phi_{max}$, and a flight altitude above the Earth's surface of $h$. This makes their distance to the barycenter of the Earth equal to $R_\oplus + h$ (see also Fig.1). The aim of such a MegaConstellation is, for example, to provide full coverage for radio or data communications up to the latitude defined by $\phi_{max}$. Note that they do not cover the entire sky.

In general, $h \ll R_\oplus$, or at least this holds for the kind of satellite close enough to the ground to exhibit a significant apparent brightness. The computations developed in this paper will use this inequality to simplify exact relationship down to those that give a first-order estimation of the dependence of various parameters on the ratio $h/R_\oplus$.

A direct consequence of the fact that $h \ll R_\oplus$ is that, although the overall number $N$ of satellites in a single layer characterized by $(h,\phi_{max})$ can be large, only a fraction of these will be visible from an observer on the ground at a given time. An even smaller proportion will be observable at zenith distances smaller than $z_{max}$ (or, equivalently, at elevations larger than $90^o-z_{max}$). In addition to these reductions, not all satellites above the zenith limit will be illuminated by the Sun at a certain specific time of night.

Astronomical observations are usually carried out at modest airmasses, to the extent that that several large facilities are built in a manner that normally cannot observe below certain elevations (Dierickx \& Gilmozzzi 1999, Ruppresch 2005, Monaco \& Snodgrass 2008, Huang 1996, Mansfield 1998, Ray 1992). Also, most large facilities now employ Active Optics (Wilson et al. 1987)  to maintain mirror figure during changes of elevation, and these systems can become unavailable when the telescope is pointed too far from the local zenith. Typically a limit of $z=60^\circ$, corresponding to 2 airmasses, applies, and a zenith angle of $z\approx 70.2^\circ$ corresponding to 3 airmasses can be considered an hard limit. Observing at higher airmasses brings the simultaneous challenges of differential refraction, higher extinction, and an increase in the intensity of telluric lines. It also causes deterioration of the capabilities of adaptive optics observations and the introduction of further difficulties when handling wide field or multiconjugated adaptive optics observations.

\begin{figure}
    \centering
    \includegraphics[width=12cm]{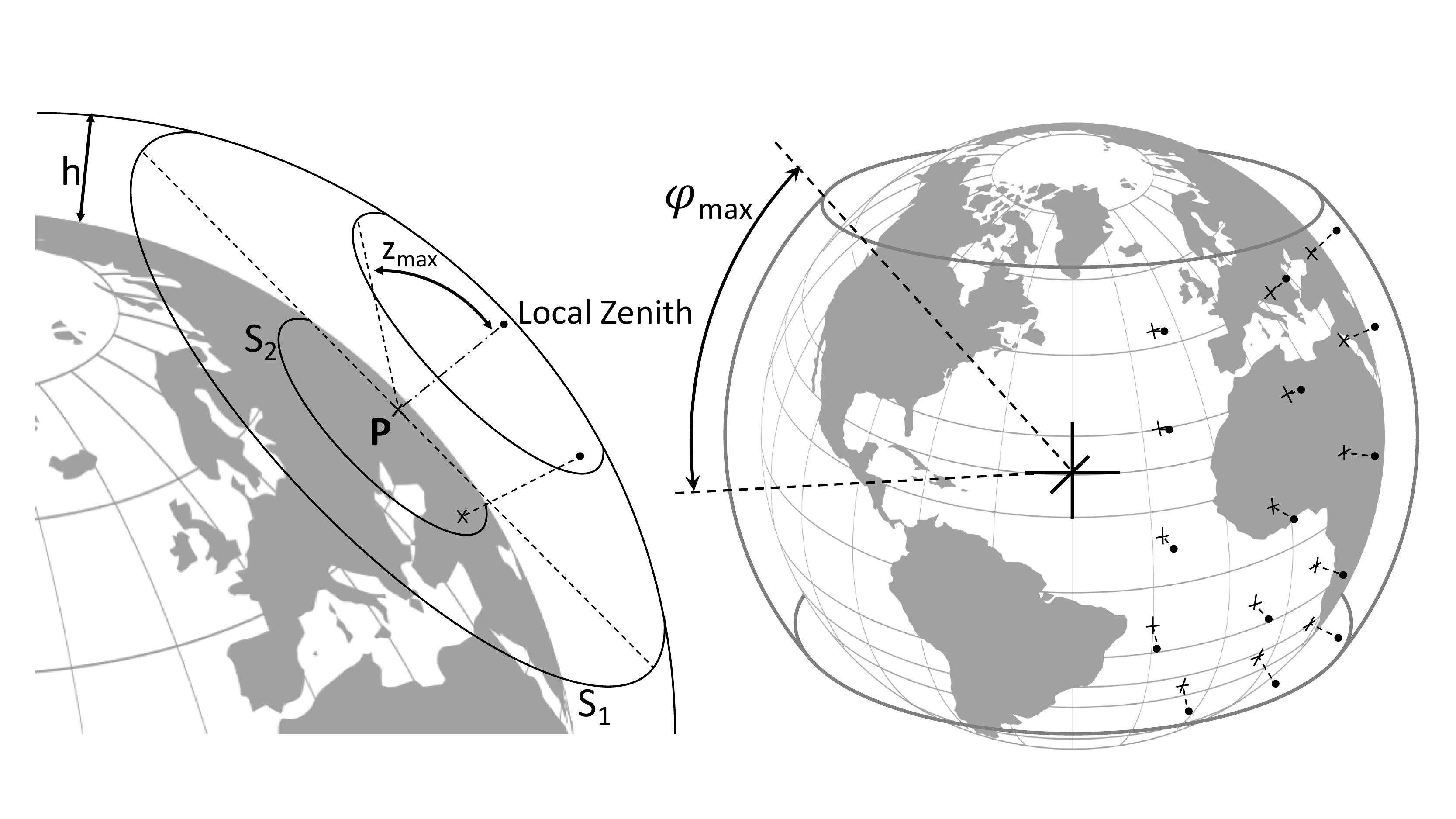}
    \caption{Around the globe, satellites flying on several circular orbits at altitude $h$, limited by a common inclination to the equator of $\phi_{max}$, are assumed to spread over such a surface, with a density given by the ratio of their number with respect to the potentially occupied surface. From a certain point $P$ on Earth, distant enough from the discontinuity at the northern and southern latitude limits, the number of satellites above the horizon can be computed using the cap's surface $S_1$. Each satellite, moreover, has a related point on Earth where the object is currently at the local zenith. Once identified, computation of the region on Earth where satellites are observed at an elevation larger than $90^\circ-z_{max}$ offers an alternative computational way so that their number can be estimated using the cap's surface $S_2$.}
    \label{blabla1}
\end{figure}

Figure 1 demonstrates that one can estimate the number of satellites observable from a certain point by computing ratios of the surfaces where the satellites fly, or the areas where such satellites are instantaneously at the local zenith. For instance, the overall surface where satellites can actually fly is given by:

\begin{equation}
    S=4\pi ( R_\oplus + h ) ^2 \sin \phi_{max}
\end{equation}

Replacing $R_\oplus +h$ with $R_\oplus$ in the above relationship yields the corresponding surface area on the Earth from which a satellite could appear at the local zenith. Given a certain point $P$ on Earth that is located distant enough to the edges defined by $\phi_{max}$, the area of the orbital sphere where all the observable satellites at a given time can lie is characterized by the cap's surface given by:

\begin{equation}
 S_1=2\pi ( R_\oplus + h ) h
\end{equation}

If $N$ is the overall population of satellites covering the orbital sphere, the number $n_1$ of these potentially visible from the observer $P$ is given by the ratio of these surfaces:

\begin{equation}
 n_1=N \eta_1 = N \frac{S_1}{S} = \frac{N}{2 \sin \phi_{max}} \frac{h}{R_\oplus + h} \approx \frac{N}{2\sin\phi_{max}} \frac{h}{R_\oplus}
\end{equation}

Note that this formula defines $\eta_1$, the ratio of the observable to overall number of satellites. Also, as anticipated, the approximation in the equation uses the fact that $h \ll R_\oplus$. 

A different and more pragmatic approach is to limit our interest to satellites which lie within an angle $z_{max}$ from zenith, as seen from $P$. We immediately recognize that, for the reasonable values of $z_{max}$ discussed before, the satellites above this elevation are a minority, since most most of them populate the more distant annulus of the cap above $P$. At these lower elevations, both the larger distance and more oblique perspective increase their apparent areal density.Using the complementary approach with points having satellites at local zenith simlilarly yelds:

\begin{equation}
 n_2 = N \eta_2 = N \frac{S_2}{4\pi R_\oplus^2 \sin \phi_{max}}   \approx \frac{N}{2\sin\phi_{max}} \frac{h^2 \tan^2 z_{max}}{R^2_\oplus}
\end{equation}

This again uses $h \ll R_\oplus$, and implicitly defines $\eta_2$. It is worth noting that, in this case, as long as $h \times \tan z_{max} \ll R_\oplus$, the first non-vanishing term is quadratic in $h/R_\oplus$, in contrast to eq.(3) where the first term is linear in $h/R_\oplus$.

Furthermore, recall that, in order to produce unwanted additional light on the focal plane, the satellites must be illuminated by the Sun. While a detailed discussion and computation can be found elsewhere (see sect. 3 of Hainaut \& Williams 2020), we note here that the angle below the horizon at which the Sun must be located to illuminate a satellite flying at altitude $h$ and crossing the local zenith is: 

\begin{equation}
    z_\odot = \arccos \frac{R_\oplus}{R_\oplus + h}
\end{equation}

If $z_\odot = 18^\circ$ (defining astronomical twilight), a satellite crossing the local zenith is just barely illuminated by the Sun at the beginning or end of astronomical night. The actual number of satellites illuminated by the Sun in these circumstances is not precisely half of that potentially visible. 
 
In winter, the situation will rapidly evolve into one in which most, if not all, of the satellites are actually not illuminated at all, especially if considering only those close enough to the local zenith to be practically observable with modern large ground based facilities.

In summer, there could be a residual region of the sky where, for the whole night, satellites are potentially illuminated. Astronomical facilities located away from the temperate zone (where, however, most of the largest facilities are located, with some notable exceptions) are potentially more affected by this. This is offset by a rapid decrease of the illuminated portion of the sky during winter, and not being disturbed by most of the MegaConstellation, if the latitude of the observatory exceeds $\phi_{max}$). A thorough discussion of this is beyond the scope of this paper. Interested readers can consult the references given in the introduction.

A summary for all these values applied to a simplified description of the StarLink MegaConstellation appears in Tab.1

\begin{table}
    \centering
    \begin{tabular}{cccccccccc}
    \hline\hline
       $h$ [km]  & $\phi_{max}$ & $N$ & $\eta_1$ &  $n_1$ & $z_{max}$ & $\eta_2$  &  $n_2$ & $z_\odot$ \\
       \hline
       $\approx${\it 340}  & $\approx${\it 53.0}$^\circ$ & {\it 25532} & {\it 3.16\%} & {\it 807} & {\it 60.0}$^\circ$ & {\it 0.34\%} & {\it 87}  & {\it 18.3}$^\circ$ \\
            &    &      &        &     & {\it 70.2}$^\circ$ & {\it 0.87\%} & {\it 222} &   \\
            \hline
       540  & $53.2^\circ$ & 1584 & 4.86\% & 77 & $60.0^\circ$ & 1.33\% & 21  & $22.8^\circ$ \\
            &      &       &        &     & $70.2^\circ$ & 3.43\% & 54 &   \\
            \hline
       550  & $53.0^\circ$   & 1584 & 4.95\% & 78 & $60.0^\circ$ & 1.39\% & 22  & $22.9^\circ$ \\
            &      &      &        &     & $70.2^\circ$ & 3.57\% & 57 &   \\
            \hline
        $\approx${\it 550} & $\approx${\it 45.0}$^\circ$ & {\it 4468} & {\it 5.59\%} & {\it 250}& {\it 60.0}$^\circ$ & {\it 1.57\%} & {\it 70} & {\it 22.9}$^\circ$ \\
                           &                             &               &  &  & {\it 70.2}$^\circ$ & {\it 4.03\%} & {\it 180} \\
                           \hline
    560     & $97.6^\circ$ & 560 & 4.06\% & 23 & $60.0^\circ$ & 1.12\% & 6 & $23.1^\circ$ \\
            &              &     & & & $70.2^\circ$ &  2.87\% &  16 & \\
            \hline
    570     & $70.0^\circ$ & 720 & 4.35\% & 31 & $60.0^\circ$ & 1.27\% & 9  & $23.3^\circ$ \\
            &    &      &        &     & $70.2^\circ$ & 3.26\% & 23 &   \\
            \hline
            $\approx${\it 1100} & $\approx${\it 53.8}$^\circ$ &  & {\it 9.09\%} & & {\it 60.0}$^\circ$ & {\it 5.49\%} &  & {\it  31.0}$^\circ$ \\
                 & & &      & & {\it 70.2}$^\circ$ & {\it 14.1\%} & & \\
       \hline\hline
    \end{tabular}
    \caption{A simplified list of the satellites in FCC Report 2020a. The italics denote a second generation group of satellites described in FCC Report 2020b, while a former layer at $h \approx 1100$km, originally included in the FCC Report 2017, is reported for comparative purposes. As all of the $z_\odot$ for the actually planned satellites are at or above $18^\circ$, a significant fraction of $n_1$ (visible in the whole sky) and $n_2$ (up to a given $z_{max}$) are still illuminated by the Sun at the beginning and the end of the astronomical night. Because of the evolving number of satellites planned in the near future, one should pay attention to the relative values of $\eta_1$ and $\eta_2$ ,rather than to the absolute numbers $n_1$ and $n_2$. Of course, the tabulated values refer only to the StarLink satellites considered here.}
    \label{tab:my_label-tab1}
\end{table}

\section{Appearance on the focal plane}

In the following, we assume the situation depicted in Fig.2, in order to derive the relationships needed to understand the impact of satellite passages in the focal plane. A satellite is crossing the local zenith and experiences constant solar illumination. $V_0$ is the apparent magnitude while crossing the zenith, and its value can be expressed in any kind of photometric system, as all of the following considerations are purely geometrical.

\begin{figure}
    \centering
    \includegraphics[width=12cm]{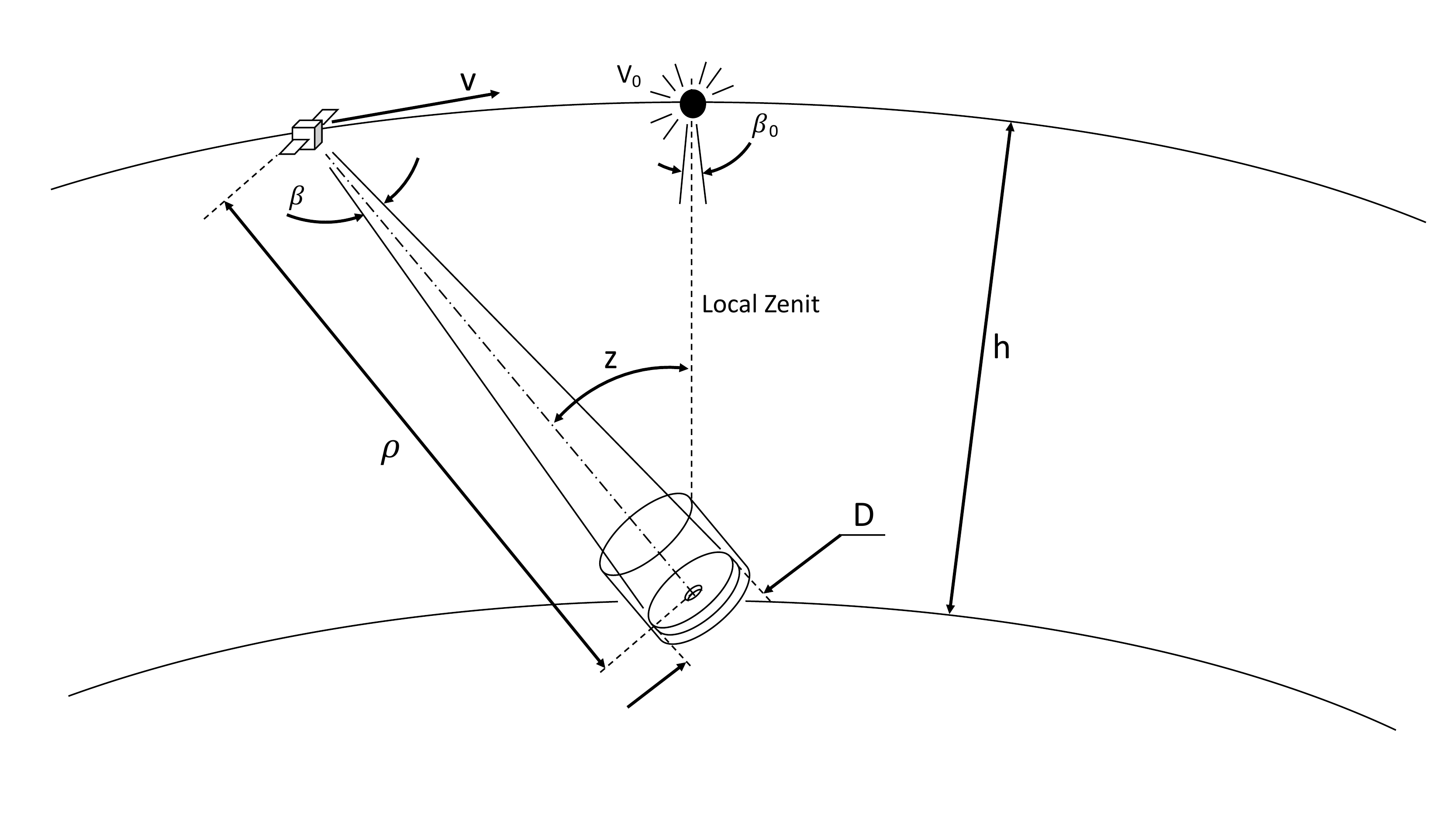}
    \caption{Definition of a number of quantities, assuming the satellite will cross the local zenith. The satellite executes its circular orbit at a relatively constant speed $v$ (neglecting Earth's rotation), and at any instant will have a range $\rho$ from the observer. At the moment of zenith transit, its integrated magnitude is $V_0$ under some kind of photometric system. Its image, produced by a telescope with diameter $D$ focused at infinity will be a defocused image of the pupil with diameter $\beta = \beta_0$ at zenith. Note that, due to varying $\rho$, the pupil image size will vary during the transit.}
    \label{blabla2}
\end{figure}

With the satellite flying at a certain constant linear speed and neglecting the relatively small effects due to the rotation of the Earth (an assumption that is very valid for the kinds of orbits discussed here by about four orders of magnitude), the range from the satellite to observer evolves with respect to the apparent zenith range as follows (and again using $h\ll R_\oplus$):

\begin{equation}
    \rho = - {R_\oplus} \cos z +\sqrt{\left(R_\oplus + h\right)^2-R_\oplus^2 \sin ^2 z } \approx \frac{h}{\cos z}
\end{equation}

while its apparent angular speed as seen from the observer (in radians per second) is (using the same approximation):

\begin{equation}
    \dot z = \frac{v}{\rho} \cdot \frac{\rho^2 +h^2+2R_\oplus h}{2(R_\oplus +h) \rho} \approx \frac{v h}{\rho^2}  
\end{equation}

These two relationships describes the evolution with time of the distance of the source and hence of its integrated brightness, due to the inverse square law. However, in order to establish the detailed surface brightness in the focal plane, we need one last parameter: the apparent angular dimension of the image.

There are three different effects that contribute to the apparent angular size of the satellite images, and each deserves at least a brief discussion:

\begin{itemize}
    \item {\bf Defocusing due to the finite distance}\\
    An astronomical telescope is of course focused at infinity. A more nearby source at a distance $\rho$ will appear as a blurred spot of size \begin{equation}
    \beta\approx D/\rho    
    \end{equation}
    when the focal plane is optically conjugated to a point at infinity. Here, $D$ is the diameter of the telescope.

    When $\beta$ is smaller than the normal image quality of the optical system, this defocus effect becomes unnoticeable. If the performance limit is given by diffraction ($\lambda/D$) or by the local seeing, one can define a minimum distance $h_{min}$ that is sometimes called the minimum depth of focus. In the case of a diffraction-limited telescope, $h_{min}$ is:
    
    \begin{equation}
        h_{min} \approx \frac{D^2}{\lambda}
    \end{equation}
    
    while for a seeing-limited imaging system, assuming $\mu$ is the seeing expressed in radians, this minimum depth of focus becomes:
    
    \begin{equation}
        h_{min} \approx \frac{D}{h}
    \end{equation}
    
     For a $D=8$m diffraction-limited telescope and $\lambda=500$nm, $h_{min}$ is about 1/3 of the Earth-Moon distance, while for a conventional seeing-limited facility, $h_{min}$ drops to about 1650km in one arcsecond seeing. This explains why no refocussing is needed when observing any class of heavenly body: they are all practically at infinity (or more distant than any reasonable $h_{min}$). This also means that, when using telescopes much smaller than the current state of the art facilities, even satellites in low Earth orbit are practically at infinity. This is not the case under investigation, however. Indeed, for the largest facilities currently operating or planned, defocus is the largest contributor to the apparent size of these satellites.

    \item {\bf Finite physical size of the satellite}\\
    An observer will experience an apparent angular size $s/\rho$ for a physical size of the satellite of $s$. However, the actual value of $s$ to use depends on the surface brightness distribution of the satellite (which can include effects from the occasional direct reflection of solar light, often called {\it flares}, depending upon the Sun direction) and on its current orientation with respect to the line of sight of the observer. A thin mast is not going to contribute significantly to the reflected flux compared to the main body of the satellite, for example, and hence will not influence the overall $s$ as intended here. 

    The maximum angular size is moreover bounded by $s/h$, meaning that, for $s=1m$, the finite size has a non-negligible maximum value of about $0.6"$, comparable to median vs. good seeing conditions. Note that the apparent angular size does not scale with the telescope aperture. For satellites in which one or two of the dimensions dominates with respect to the others, the apparent value of $s$ can vary wildly due to its relative attitude. Because of the difficulties in predicting the intrinsic angular size, we neglect its influence in the following discussion. This will result in an underestimation of the actual apparent angular size of the satellites. We also note that the angular size is of the same order of magnitude as the uncertainty in the estimates of  {\it average} seeing as experienced at good observing sites, In this way, it can be included in the uncertainty in the final result.

    \item {\bf Seeing}\\
    With the exception of Adaptive Optics assisted diffraction-limited, any image will be strongly affected by overall atmospheric disturbance. Note that the ``cone effect'', due to the fact that the light from the satellite experiences a different volume of atmosphere than an object at infinity, is likely to make a negligible, if measurable at all, difference in the final image size. However, speckles could makes visible effects on the trails, as they are  projected onto a specific position on the focal plane for an extremely short time, potentially freezing eventually such variations. In the following, I assume a nominal seeing of $\mu"=1"$. In all the cases described in this paper, in other words, where the apparent size is dominated by the defocus, variations in seeing have little effect on the final result.

\end{itemize}

The defocused image of an unresolved source is a small image of the pupil of the telescope, often characterized by a circular shape with a small central obstruction. The seeing (and the possible effects of the finite size) will makes the edges blurred and the central obstruction less noticeable or even indiscernible under some conditions.
In the following, I describe the shape of such sources in the focal plane as a circular disk of diameter $\theta"$ whose size is of the order of 

\begin{equation}
    \theta" \approx \sqrt{\beta"^2+\mu"^2}
\end{equation}

where $\beta" \approx 206265 \times \beta$ to express it in arcseconds. These round spots will travel on the focal plane at the apparent speed given by $\dot z$, and one can define a maximum travel time (measured along the diameter) of this bundle of light in the focal plane:

\begin{equation}
    \Delta t \approx \frac{\theta"}{\dot z"}
\end{equation}

At the same time, the overall brightness of the sources will evolve with range following an inverse square law projected on the magnitude scale:

\begin{equation}
    V = V_0 + 5 \log_{10} \left(\frac{\rho}{h}\right)
\end{equation}

Neglecting the effect of seeing, these two effects cancel each other such that the surface brightness remains constant.

\begin{figure}
    \centering
    \includegraphics[width=12cm]{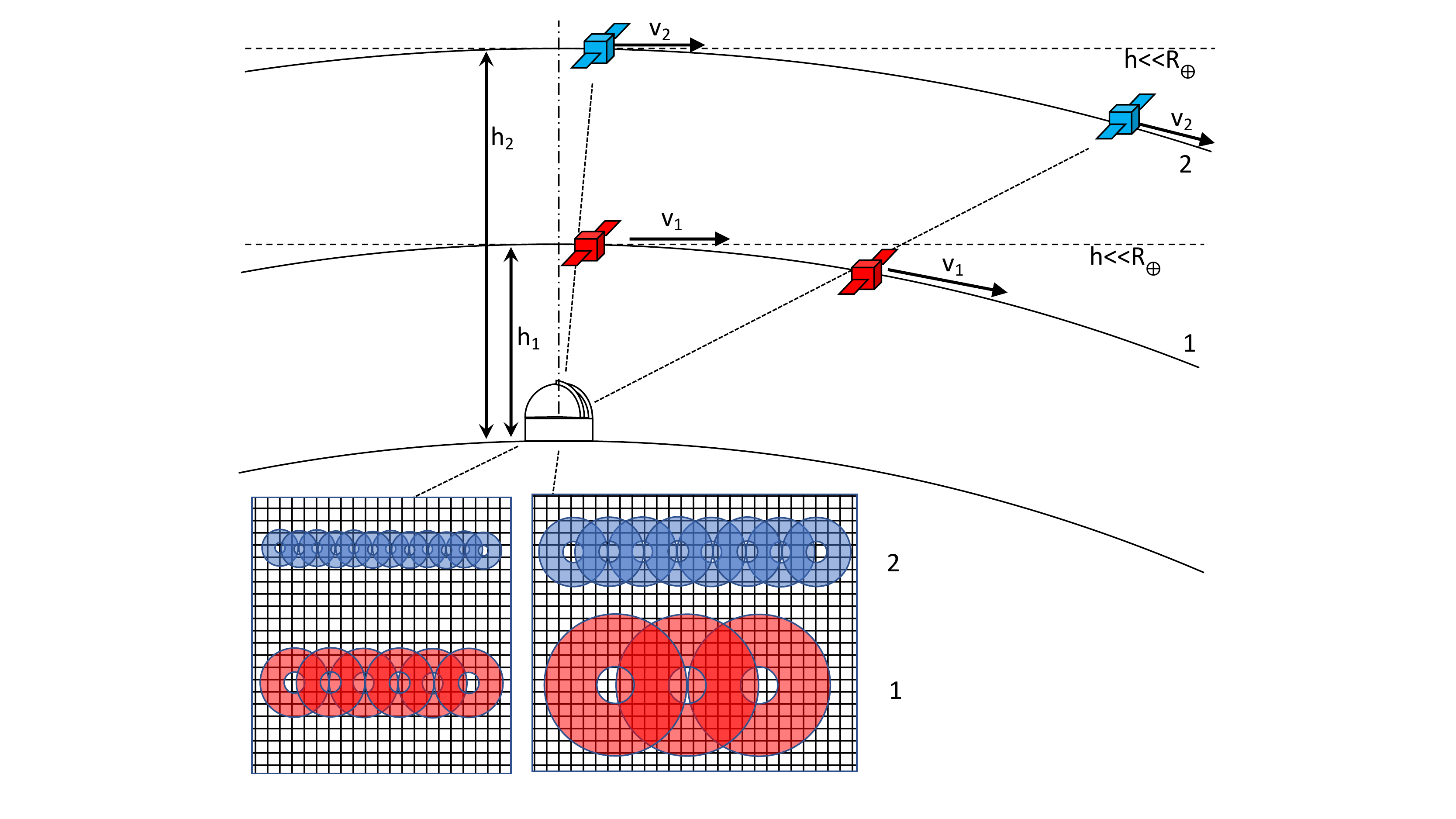}
    \caption{Let us consider the same kind of satellite in term of actual brightness when illuminated by the Sun, placed into two circular orbits $1$ and $2$. As long as the defocus term is dominant with respect to others, such as the seeing, the apparent size on the focal plane will scale along with the overall brightness, making the surface brightness invariant. Furthermore, if the satellite is observed along a portion of its orbit where the approximation $h\ll R_\odot$ holds, the only variation in the brightness at different elevation is due to the difference in absolute tangential velocities $v_1$ and $v_2$.}
    \label{fig:my_label-xxx}
\end{figure}

This means that, instantaneously, the light is spread onto a non-negligible area, that is usually significantly larger than the seeing. This is especially true for the largest facilities, leading to an instantaneous surface brightness in equivalent magnitude per square arcsecond given approximately by:

\begin{equation}
    V_{\boxvoid"} = V + 2.5 \log_{10} \left( \frac{\pi}{4} \theta"^2 \right) \approx V -0.26+5 \log_{10} \theta"
\end{equation}

When the defocusing effect becomes smaller or negligible with respect to the seeing $\mu"$ due to larger range $\rho$, the brightness of the satellites will consistently diminish following the inverse square law.

However, for an exposure time $t \gg \Delta t$, this surface brightness illuminates the focal plane for only a fraction of the exposure time and it will be recorded as proportionally diluted with respect to the other fixed sources (including the sky background). At the end of the exposure, the satellite will produce a trail whose width is of the order of $\theta"$ and with an equivalent surface brightness given by using the appropriate $\Delta t /t$ factor (adjusted to the magnitude scale) in the last equation:

\begin{equation}
    V_{\boxvoid"t} = V_{\boxvoid"} +2.5 \log_{10} \left( \frac{\Delta t}{t} \right)
\end{equation}

It is interesting that a number of approximations overestimate the effects. For example, the travel time is only applicable to the central part of the trail. Specifically, the brightness will be the calculated value only at its center and the image will fade toward the edges. The calculations also assume that the defocused spot is a uniform circular disk. For the typical image size of the order of a few arcsec appropriate to $D=4..8$m class telescopes, this is a reasonable assumption, because the central obstruction will likely be washed out by the smearing effect of the seeing and from the size of the satellite (this last parameter being ignored here). Table 2 presents derived values for a range of reasonable parameters.

Figure 3 demonstrates that, for a given kind of satellite, the apparent instantaneous surface brightness is unchanged when its distance $\rho$ evolves with time. The same concept applies when the satellite occupies an orbit with a different $h$. For example, in the situation depicted in Fig.3, the same satellite located at twice the distance from the observer will have an integrated brightness four times smaller. However, the image is produced in the focal plane through a defocusing two times smaller, thus retaining the same surface brightness. In a longer exposure, an angular speed exactly half as great as that of the closer satellite will make the diluted surface brightness equivalent again. Note that, for the quantities considered here, the effect of the difference in actual velocity is  less than the effect of the actual defocus size, including the seeing. In other words, to a first approximation, the surface brightness of trails does not depend upon either the elevation $z$ or the orbit height $h$ for the kind of low orbit considered here. And, at least for the current generation of existing large telescopes, the brightness is altered more by the seeing-induced enlargement of the defocused disk than by the variation in speed due to the different orbit. Comparing orbits at $h_1=340$km and $h_2=550$km, the latter has a variation in tangential speed  of about 1.6\%, while for a $D=8$m telescope, seeing of $\mu"=1"$ will change the corresponding defocus size from $\beta"$ to $\theta"$ by more than about 2\% (taking the most favourable case of a $D=8$m telescope and an $h=340$km satellite). This means that the effect of the differing tangential orbital speeds is (barely) relevant only for a much larger aperture (or for much more distant satellites -- however, these will appear significantly fainter).

Figure 4 shows the integrated magnitude plotted vs. elevation for the two case studies used in the remainder of this paper\footnote{Figs. 4, 5 and 6 have been generated using an IDL procedure. A version comprehensive of all the informations of these plots with most of the parameters adjustable via keywords is available at: shorturl.at/vDIP1}. The zero point has been scaled from theDarkSat measurements of Tregloan-Reed et al. 2020 and converted into visual magnitudes using a Sun like spectrum as reference (and thereby implicitly assuming that the satellite is {\it gray}) following Fukugita et al. 1996).

Figure 5 plots the instantaneous surface magnitude vs zenith distance for six different combinations of satellite orbits and telescope apertures, while Figure 6 presents the dilution effect for exposures of 1 second, 1 minute and 1 hour. Note that in both Figure 5 and 6, the scale on the left is tuned to the DarkSat parameters, while the right-hand scale is computed assuming $V_0=0$ for a $h=340$km orbit. This allows easy scaling to any preferred satellite. 

Note also that, with transit times across the focal plane on the order of milliseconds, a one second exposure fully qualifies as a {\it long} exposure.

\begin{table}
    \centering
    \begin{tabular}{ccccccccc}
    \hline\hline
        $h$ [km] & $v$ [km/s] &$\dot z_0$ & $V_0$ & $D$ [m] & $\beta_0"$ & $V_{\boxvoid"}$ &  $\Delta t$ [mSec.] & $t_{min}$ \\ 
        \hline
        340 & 7.70 & $1.30^\circ$/sec  & 5.02 &  4 & 2.4" & 6.83 &  0.56 & 10'55"\\
            &      &  & & 8 & 4.9" & 8.25 &   1.07 & 5'38" \\
            &      & & & 40 &  24.3" &  11.69 &                 5.20  & 1'09"\\
            \hline
        550 & 7.58 & $0.79^\circ$/sec  & 6.06 &  4 & 1.5" & 7.07 &  0.63 & 9'51"\\
            &      & & & 8 & 3.0" & 8.30 &   1.11 & 5'35" \\
            &      & & & 40 & 15.0"  &    11.68 &      5.28 & 1'11"\\
            \hline
    1100    &  7.30  &  $0.39^\circ$/sec      & 7.57 &  4  &     0.75"    & 7.76 &   0.87 &  7'12"  \\
            &     & & & 8   & 1.50"        & 8.59 &           1.27 &   4'54"\\
            &     & & & 40 & 7.5"   &  11.70 &         5.33 &   1'10"  \\
        \hline \hline
    \end{tabular}
    \caption{Parameters for three altitudes $h$ of the same kind of sunlit satellite, consistent with DarkSat brightness and flying on circular orbits. The second through fourth columns list the tangential velocity $v$, the apparent angular speed $\dot z_0$, and the zenith brightness $V_0$. For each of the three telescope diameters $D$, subsequent columns list the corresponding apparent defocus size $\beta_0"$, approximate instantaneous surface brightness of the spot $V_{\boxvoid"}$ in magnitudes per square arcsecond, and the travel time $\Delta t$ to cross its own diameter. The latter two values include the effect of seeing assuming $\mu'=1'$. Finally, the last column shows the exposure time for which the trail surface brightness is equivalent to a $V_{\boxvoid"}=22.0$ (see for instance Patat 2008 and references therein for a discussion on the dark sky brightness).}
    \label{tab:my_label2}
\end{table}

A number of derived quantities can be easily worked out in order to assess the effects of these trails. For example, one could define the minimum exposure time for which the trails are comparable to, or a fraction of, the natural sky brightness. Table 2 lists some of these quantities to give a rough order of magnitude of these numbers.

\begin{figure}
    \centering
 \includegraphics[width=12cm]{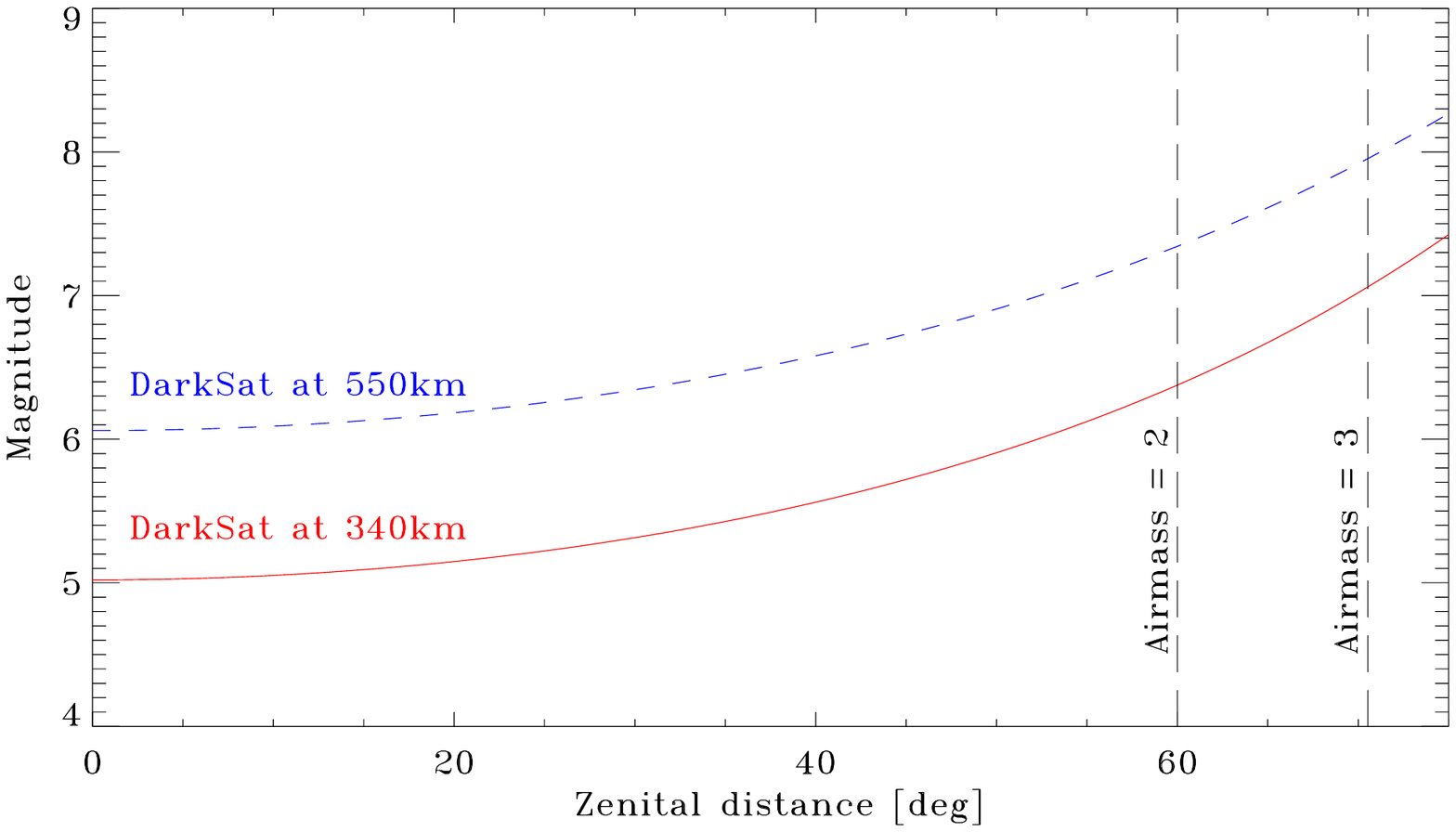}
    \caption{The same satellite, DarkSat in this example and in the following figures, placed at two different altitudes in a circular orbit, will vary in brightness with elevation, due to the inverse square law.}
    \label{NewFig3}
\end{figure}

\begin{figure}
\centering
\includegraphics[width=12cm]{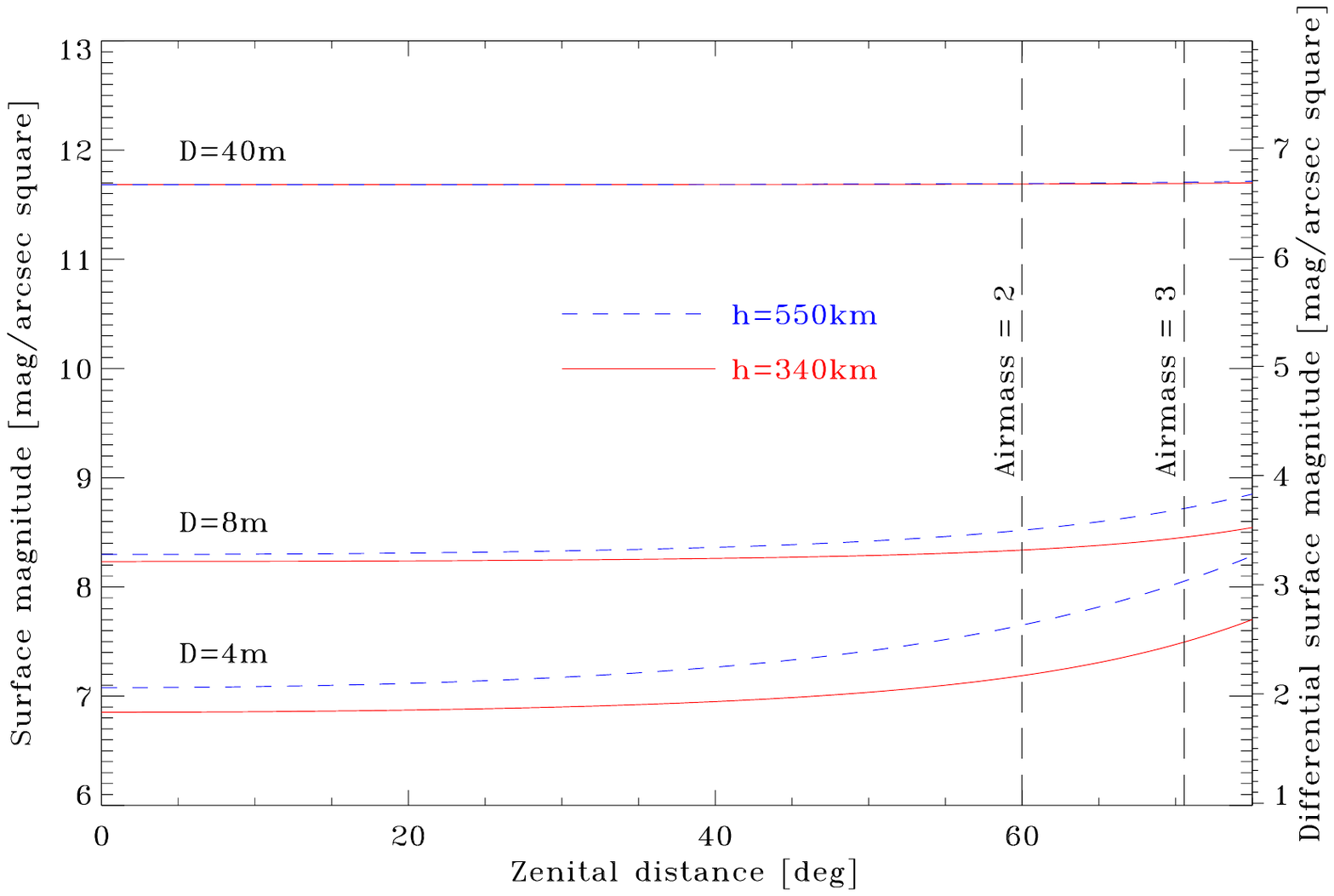}
\caption{The same DarkSat satellite at different altitudes will exhibit approximately constant instantaneous surface brightness at the local zenith for a specific telescope aperture $D$, the only difference being due to the influence of the finite effect in the size of the defocussed spot introduced by the local seeing. Because this small variation in the brightness is due to the combination of the telescope diameter and the actual distance from the observer, this effect is larger for smaller apertures and large zenith distances. The vertical scale on the left applies to the DarkSat case, while that on the right can be scaled to a particular case by adding the magnitude at zenith transit $V_0$ at $h=340$km. In other words, when the effect of the seeing starts to dominate with respect to the defocus, the instantaneous surface brightness will approach the inverse square law depicted in the previous figure.}
\label{NewFig4}
\end{figure}

\begin{figure}
\centering
\includegraphics[width=12cm]{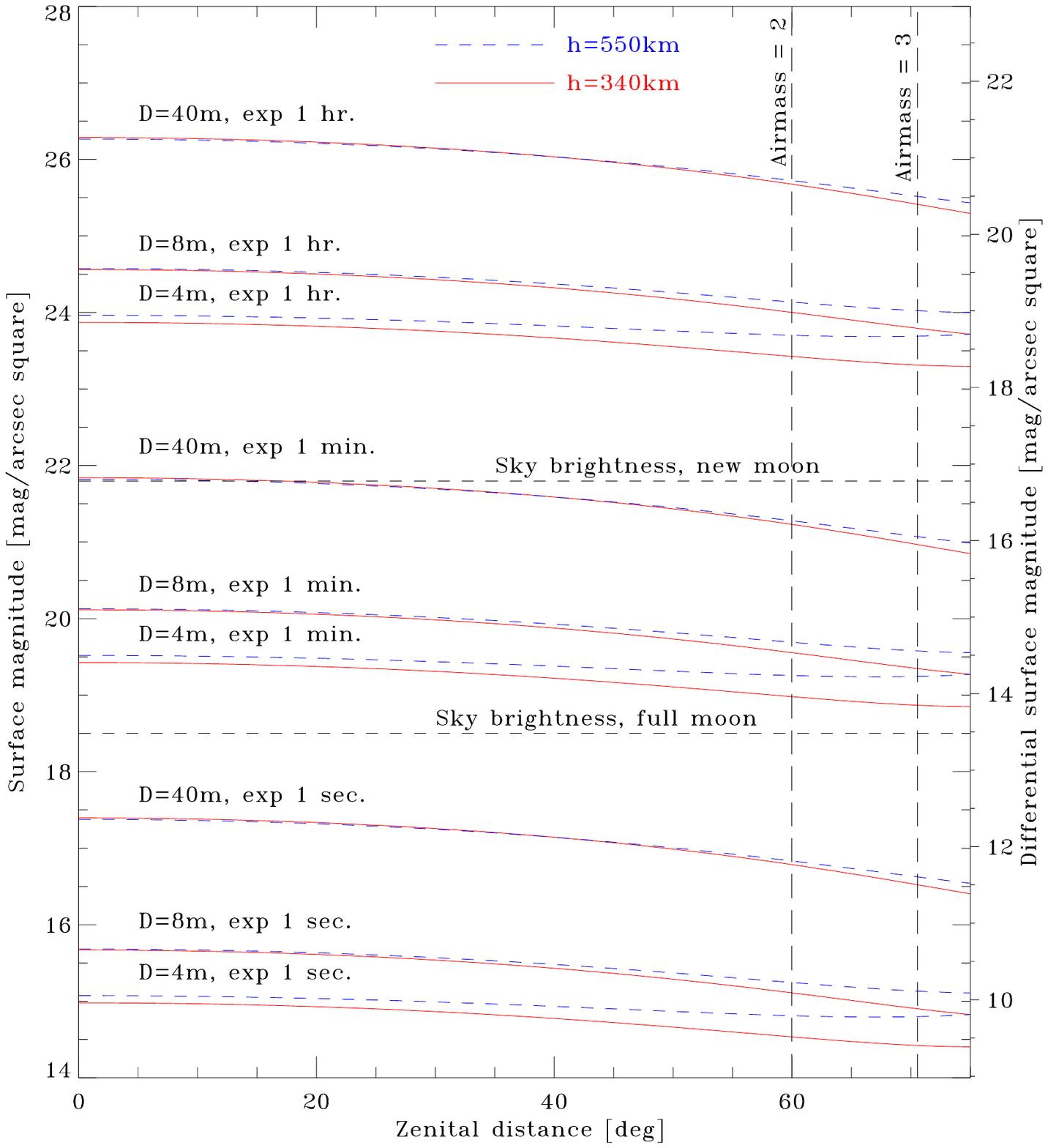}
\caption{Because of the angular motion on the sky, the trails will appear diluted when recorded on an exposure longer than the travel time of the defocused image on the focal plane. As this last value is of the order of milliseconds, a one second exposure is already enough to make the effect evident. The relative effect of the seeing of the defocused disk size (as shown by the differences between the solid/red vs. dashed/blue lines) is larger for the $D=4$ and $D=8$m telescopes case, while it is slightly overcome by the smaller absolute tangential speed of the satellites only in the case of the $D=40$m telescope. In the latter instance, the lowest altitude satellite is fainter than the higher one because of its faster absolute movement. The effect of the circular orbit becomes evident when observing at larger elevations, where the satellites is closer that would expected by a $h\ll R_\odot$ approximation. This makes the actual apparent integrated brightness of the trails somewhat larger, especially for the lowest orbits.}
\label{NewFig5}
\end{figure}

One can note, furthermore, that at larger distance (and hence zenith distance) the increased blurring due to the seeing becomes more relevant, especially for smaller apertures, and makes the surface brightness slightly fainter. However, because of geometrical effects, the apparent speed of the image compared to the diameter of the defcoused disk is lower, making the equivalent surface magnitude for a long exposure brighter as the target approaches the horizon. Within a Field of View of even a few degrees, the surface magnitude rarely change by more than a small fraction of magnitude. It is also worth noting that the dilution of this surface brightness, even for a one second exposure, makes the equivalent brightness in the region comparable to one of the brightest planetary nebulae observable from Earth.

\section{Conclusions}

The fraction of satellites above the horizon at a given point on the Earth is approximately $1/2 \times h/R_\oplus$, while the the fraction at less than 2 airmasses is of the order of $3/2 \times (h/R_\oplus)^2$. Despite the impressive  number of satellites involved in MegaConstellations, about 1\% or less (see $\eta_2$ in Tab.1) are potentially in sight over a realistic range of elevations for astronomical observations. A larger fraction appears, much fainter, in a strip above the horizon.  At the beginning or end of the astronomical night, several satellites will be illuminated, and the ones at the lower orbit will still be barely illuminated at  zenith. The higher ones will continue to be illuminated, even if crossing the local zenith, until the Sun falls lower by several degrees. The surface brightness of those illuminated by the sun is, however, more or less constant for a specific kind of satellite and, for the case considered in this paper, the surface brightness is fainter than that of the planet Uranus (Zombeck, 2007) on a $D=8m$ telescope. As the defocused spots, a few arcsec in size, travel across the focal plane at  several thousand arcseconds per second, the flux is diluted over the whole frame, even in a 1 second exposure, down to a level similar to that of the peaks of the brightest planetary nebula. This casts doubt on the possibility that such trails could ruin the whole frame by some sort of blooming effect, as pointed out by Hainaut \& Williams 2020 and Tyson 2020, for whom the experiments carried out with subarcsec width trails led to this worry.

Longer exposures make the dilution effect such that, in one hour equivalent exposure (even stacking shorter exposures), it could be hard to actually detect the satellites (although this is not necessarily a positive remark, as objects in the trails could, by consequence, be erroneously photometrically measured). The altitude of the same kind of satellite will make no difference unless their distance is outside of the depth of focus of the telescope conjugated to infinity. Beyond this distance, the actual surface brightness will drop, making the overall effect less relevant, although they could be illuminated even in some significant portion of the astronomical night.

It is worth noting that both the statistical and numerical analyses carried out in previous work assume that no one is  going to actively avoid having these trails fall into the field of view of their observations. In principle, the location of satellites and hence their trails can be predicted with great accuracy, assuming the ephemerides are known. However, scheduling schemes for satellite avoidance, which could probably work well for small Field of View telescopes, can have limited or no effects for larger field of view telescopes as discussed in section 2 of Tyson et al. 2020.

Observations at dawn and dusk can be affected, as pointed out by others (see for instance McDowell 2020), but a fair comparison should recognize that the natural sky brightness is much greater at that time, and the comparison should take this into account. As the satellite brightness changes rapidly with time and direction, this aspect deserves a specific detailed computation.

As mentioned earlier, all of these considerations change dramatically with smaller apertures, since the satellites will be basically unresolved and appear in focus for any relatively small optics, including the naked eye. This will make these satellites visible or barely visible to the naked eye, and will significantly affect astrophotographers.  

\section*{Acknowledgements}

Thanks are due to an anonymous Referee for suggestions that led to a significant improvement of the paper, and to Tom Herbst of MPIA, Germany, for improving the readability of the final work.

\section*{References}

\noindent
Dierickx P. \& Gilmozzi R. 2000 in ESO conf. proc. 57 “Proceedings of the Bäckaskog Workshop on Extremely Large Telescopes” edited by Andersen T., Ardeberg A., Gilmozzi R. (Lund, Sweden, ESO)  43



\noindent
FCC Report 2017, SAT-LOA-20170301-00027, narrative downloadable through: https://fcc.report/IBFS/SAT-LOA-20170301-00027/1190018.pdf

\noindent
FCC Report 2020a, SAT-MOD-20200417-00037, narrative downloadable through: https://fcc.report/IBFS/SAT-MOD-20200417-00037/2274315.pdf

\noindent
FCC Report 2020b, SAT-LOA-20200526-00055, narrative downloadable through: https://fcc.report/IBFS/SAT-LOA-20200526-00055/2378669.pdf

\noindent
Fukugita M., Ichikawa T., Gunn J.E. et al. 1996, AJ 111, 1748

\noindent
Hainaut O.R. \& Williams A.P. 2020 A\&A 636, A121

\noindent
Huang E.W. 1996 SPIE proc. 2871, 291

\noindent
Mansfield A. 1998 SPIE proc. 3352, 112

\noindent
McDowell J.C.  2020 ApJL 892, L36

\noindent
Monaco L. \& Snodgrass C. 2008 “EFOCS2 User’s Manual” LSO-MAN-ESO-36100-0004

\noindent
Patat F. 2008, A\&A 481, 575

\noindent
Ray F.B. 1992 in ESO conf. proc. 42 “Progress in Telescope and Instrumentation Technologies” edited by Ulrich M.-H. (Garching, Germany, ESO)  91

\noindent
Rupprecht  G. 2005, “Requirement for Scientific Instruments on the VLT Unit Telescopes” VLT-SPE-ESO-1000-2733

\noindent
Tyson J.A. 2020 meeting material \#03 from: https://www.nationalacademies.org/event/04-27-2020/decadal-survey-on-astronomy-and-astrophysics-2020-astro2020-light-pollution-rfi-meeting

\noindent
Tyson J.A., Ivezic, Z., Bradshaw A. et al. 2020 arXiv:2006.12417v1 

\noindent
Tregloan-Reed J., Otarolas A., Ortiz E. et al. 2020 A\&A 637, L1

\noindent
Wilson R.N., Franza F. \& Noethe L. 1987 J. of Mod. Opt. 34, 485

\noindent
Zombeck V.M. 2007 Handbook of Space Astronomy and Astrophysics, Cambridge Press, ISBN-13 978-0-521-78242-5

\vspace{1cm}
\noindent
The current version of the manuscript is consistent with the one accepted to appears in Pubblication of the Astronomical Society of the Pacific

\end{document}